\input phyzzx.tex
\normalbaselineskip = 16pt plus 0.2pt minus 0.1pt


\def\p{\pi}
\def\s{\sigma}
\def\t{\tau}


\REF\Polchinski{J. Polchinski, Phys. Rev. Lett. {\bf 75} (1995) 4724, 
hep-th/9510017}
\REF\GS{M.B. Green and J.H. Schwarz, Phys. Lett. {\bf 136B} (1984) 367.}
\REF\GG{M.B. Green and M. Gutperle, Nucl. Phys. {\bf B476} (1996) 484, 
hep-th/9604091}
\REF\EMM{K. Ezawa, Y. Matsuo and K. Murakami,
Phys. Rev. {\bf D57}, 5118 (1998),
hep-th/9707200.}
\REF\DLP{J. Dai, R.G. Leigh and J. Polchinski, Mod. Phys. Lett. 
{\bf A4} (1989) 2073.}
\REF\HW{A. Hanany and E. Witten, Nucl. Phys. {\bf B492} (1997) 152, 
hep-th/9611230.}
\REF\BN{L. Brink and H.B. Nelson, Phys. Lett. {\bf B45} (1973) 332.}
\REF\Witten{E. Witten, Nucl. Phys. {\bf B460} (1996) 335, hep-th/9510135.}
\REF\Green{M.B. Green, Phys. Lett. {\bf B329} (1994) 435, hep-th/9403040.}
\REF\PW{J. Polchinski and E. Witten, Nucl. Phys. {\bf B460} (1996) 525, 
hep-th/9510169.}
\REF\BSV{M. Bershadsky, V. Sadov and C. Vafa, 
Nucl. Phys. {\bf B463} (1996) 398, hep-th/9510225.}

\pubnum={KCL-TH-99-17\cr hep-th/9905029}
\date{5 May 1999}

\titlepage

\title{\bf D-Branes in the Green-Schwarz Formalism}

\centerline{N.D. Lambert}

\centerline{and}

\centerline{P.C. West\foot{lambert, pwest@mth.kcl.ac.uk}}
\address{Department of Mathematics\break
         King's College, London\break
         England\break
         WC2R 2LS\break
         }

\abstract
We give a basic account of  
supersymmetric open strings and D-branes
using the Green-Schwarz formalism, obtaining 
a manifestly spacetime supersymmetric description
of their spectrum. 
In addition we discuss a mechanism whereby some of the
D-brane states are projected out and which can lead to chiral quantum field
theories on the brane.

\eject


In the past four years it has become apparent that 
D-branes are of central importance in our understanding
of string theory [\Polchinski]. An important feature of D-branes is that 
they give rise to non-Abelian vector multiplets in a variety of 
dimensions less than ten. This has led to a new era in the study of
Yang-Mills theories with many results. A key step in these studies is
to determine the spectrum of an intersecting brane configuration
from the various D-branes within it.  
In this paper we would like to provide
a study of D-branes in a manifestly spacetime supersymmetric form,
using the Green-Schwarz formalism [\GS]. While there have been detailed 
studies focusing on spacetime supersymmetry [\GG], to the best of our 
knowledge, a
basic treatment of open strings with Dirichlet boundary conditions 
using the Green-Schwarz formalism has not appeared (see [\EMM] for a
similar disscussion of open supermembranes). This provides a 
simple and elegant
description of the spectrum which we hope will be useful when studying
brane configurations and quantum field theories, particularly when studying
more phenomenological applications. We also note that a
D-brane can be defined as an open string  that
preserves a given spacetime supersymmetry. It then follows that its states only
propagate in $p+1$ dimensions planes where the string end points must lie.
We will also
describe a mechanism whereby some states  can be projected out and which can
lead to chiral fermions.

We now review the essential aspects of the Green-Schwarz
formulation of the superstring that we will require [\GS].
This 
consists of the fields
$X^{\mu }(\tau,\sigma ) ; \mu  = 0,1,\ldots ,D-1$ and
the Grassmann odd fields
$\theta ^{i}(\tau,\sigma )$ which is a
spacetime spinor with an internal index i = 1,2.
The fields $\theta^i$ also carry a spinor index that we suppress and which runs
over the dimension of the Clifford algebra. The dimension of this algebra 
varies depending on the
spacetime dimension $D$ and the type of spinor i.e. Majorana or
Majorana-Weyl.  We 
use the mostly plus signature and define the Majorana conjugate 
of a spinor by 
$\bar \theta = \theta^TC$. Here $C$ is the charge conjugation
matrix defined through the relation $(\gamma^{\mu})^T = -C\gamma^{\mu}C^{-1}$
and satisfies $C^T=-\eta C$ with $\eta=+1$ for $D=2,4\ {\rm mod} 8$ or 
$\eta=-1$ for $D=0,6\ {\rm mod}8$. 
We will mainly be interested in ten dimensions
where the spinors are Majorana-Weyl and with either the same chirality
(IIB) or opposite chirality (IIA). 

The string has the action
$$
S = S_{1}+ S_{2}\ .
\eqn\action
$$
Here the first term is given by
$$
S_{1} = - {1\over 2\pi } \int d\s d\t  \sqrt{-g} g^{\alpha \beta }
\Pi ^{\mu }_{\alpha } \Pi ^{\nu }_{\beta } \eta _{\mu \nu }\ ,
$$
where
$$ 
\Pi ^{\mu }_{\alpha }
= \partial _{\alpha }X^{\mu }
- i \bar{\theta }^{j} \gamma ^{\mu }\partial _{\alpha }\theta ^{j}\ ,
$$
and the second term has the form
$$
S_{2} = {1\over \pi } \int d\s d\t  \epsilon ^{\alpha \beta }
\{-i \partial _{\alpha }X^{\mu }(\bar{\theta }^{1}\gamma ^{\nu }
\partial _{\beta }\theta ^{1}
- \bar{\theta }^{2}\gamma ^{\nu }\partial _{\beta }\theta ^{2})
+ \bar{\theta }^{1}\gamma ^{\mu }\partial _{\alpha }\theta ^{1}
\bar{\theta }^{2}\gamma^{\nu }\partial_{\beta }\theta ^{2}\}
\eta _{\mu \nu }\ .
$$
The action is invariant under the spacetime  supersymmetry
transformations
$$\eqalign{ 
\delta X^{\mu } &= i\bar{\epsilon }^{j}\gamma ^{\mu }\theta ^{j}\ ,\cr
 \delta \theta ^{i} &= \epsilon ^{i}\ ,\cr}
\eqn\susy
$$
where $\epsilon ^{i} , i = 1, 2$ is a constant
spinor of the same chirality as
$\theta ^{i}$. It is also invariant under 
the $\kappa ^{\alpha i}$-transformations
$$\eqalign{
\delta X^{\mu }
&= i \bar{\theta }^{j} \gamma ^{\mu } \delta \theta ^{j}\ ,\cr
\delta \theta ^{i}
&= 2i \gamma ^{\mu } \Pi _{\mu \alpha } \kappa ^{i\alpha }\ .\cr}
\eqn\ksym
$$
The spinorial parameter $\kappa ^{\alpha i}$ is
subject to the projection conditions
$$
\kappa ^{1\alpha } = P^{\alpha }_{-\beta } \kappa ^{1\beta }\quad
 \kappa ^{2\alpha } = P^{\alpha }_{+\beta } \kappa^{2\beta }\ ,
$$
where
$$
P^{\alpha }_{\pm  \beta }
= {1\over 2} (\delta ^{\alpha }_{\beta } \pm
{\epsilon ^{\alpha \delta }g_{\delta \beta }\over
\sqrt{-g}})\ .
$$
The equations of motion for the action are easily found to be
$$\eqalign{
\Pi _{\alpha }\Pi _{\beta } - {1\over 2} g_{\alpha \beta } g^{\gamma \delta }
\Pi _{\gamma }\Pi _{\delta } &= 0\ ,\cr
\gamma ^{\mu } \Pi _{\mu \alpha } P^{\alpha }_{-\beta } g^{\beta \delta }
\partial _{\delta } \theta ^{1} = 0 ,\quad
 \gamma ^{\mu } \Pi _{\mu \alpha } P^{\alpha }_{+\beta} g^{\beta \delta }
\partial _{\delta } \theta ^{2} &= 0\ ,\cr
\partial _{\alpha }\{\sqrt{-g} g^{\alpha \beta }\partial _{\beta } X^{\mu }
- 2i P^{\ \alpha } _{-\ \beta } g^{\beta \delta } \bar{\theta }^{1}
\gamma ^{\mu }\partial _{\delta }\theta ^{1}
- 2i P^{\ \alpha }_{+\ \beta } g^{\beta \delta }
\bar{\theta }^{2} \gamma ^{\mu }\partial _{\delta }\theta ^{2}\} &= 0\ .}
\eqn\fulleqofm
$$

In deriving the above equations of motion  we did not discuss the
boundary terms that arise when we vary the action \action.   
Due to locality we must  adopt  boundary conditions such
that the boundary terms vanish separately at each end point. This leads to the
equation
$$
\int d\tau
\left(\delta
X^\mu{\delta S\over\delta(\partial_\s X^\mu)}
+\delta\theta^i{\delta S\over\delta
(\partial_\s\theta^i)}\right)=0\ ,
\eqn\boundary
$$
at $\sigma=0$ and $\sigma=\pi$.
Choosing the gauge  $g_{\alpha
\beta}=\eta_{\alpha
\beta}e^\phi$  we find that the above boundary
term vanishes if
$$
\eqalign{
\delta X^\mu &(\Pi_{\s\mu}+i(\bar\theta^1 \gamma_\mu\partial_\s\theta^1
-\bar\theta^2\gamma_\mu\partial_\s\theta^2))
+i\Pi_{1\mu}(\bar\theta^1\gamma^\mu\delta\theta^1 
+\bar\theta^2\gamma^\mu\delta\theta^2)\cr
&\quad-i\partial_\t X^\mu(\bar\theta^1\gamma_\mu\delta\theta^1
-\bar\theta^2\gamma_\mu\delta\theta^2)
+\bar\theta^1\gamma^\mu\partial_\t\theta^1\bar\theta^2\gamma^\nu\delta\theta^2
-\bar\theta^1\gamma^\mu\delta\theta^1
\bar\theta^2\gamma_\mu\partial_\t\theta^2\cr&=0\ ,}
\eqn\boundarytwo
$$
at the boundaries $\sigma=0$ and $\sigma=\pi$.

Analysing the first term in \boundary\ shows that 
ends of the string obey Neumann (N) or Dirichlet (D) 
boundary conditions [\DLP]. Let us denote
the directions that obey N and D boundary conditions at $\sigma=0$ by
$N_0$ and $D_0$ respectively and similarly at the other end at
$\sigma =\pi$ by $N_\pi$ and $D_\pi$. The D-brane the string is
attached to at $\sigma=0$ then has its longitudinal directions in
$\mu\ \in \ N_0$.
Since
$\delta X^\mu$ vanishes in the directions  that have D boundary
conditions, the boundary
terms containing $\delta X^\mu$ will vanish if
at $\sigma=0$ if
$$
(\bar\theta^1\gamma^\mu\partial_\t\theta^1
-\bar\theta^2\gamma^\mu\partial_\t\theta^2)
+(\bar\theta^1\gamma^\mu\partial_\s\theta^1
+\bar\theta^2\gamma^\mu\partial_\s\theta^2)=0,\ \mu\ \in \ N_0,
$$
with a similar condition at the other end of the string at
$\sigma=\pi$.  Below we  will focus on the conditions at
$\sigma=0$, but those for $\sigma=\pi$ are essentially identical.

Examining the $\delta\theta^i$ variations we find that
at $\sigma=0$
$$
\eqalign{
\bar\theta^1\gamma^\mu\delta\theta^1
+\bar\theta^2\gamma^{\mu}\delta\theta^2&=0, \ \mu\ \in \ D_0,\cr
\bar\theta^1\gamma^\mu\delta\theta^1
-\bar\theta^2\gamma^\mu\delta\theta^2&=0,\ \mu\ \in \ N_0.\cr}
$$
The difference in sign corresponds to fact that
$\partial_\s X^\mu$ and
$\partial_\t X^\mu$ are undetermined
in the D and N directions respectively. The higher order
terms in $(\theta)^3$ must vanish separately. We must place a
condition on $\theta^i$ of the form $\theta^2=P_0\theta^1$. Since
$\theta^1$ and $\theta^2$ are Majorana spinors this implies that
$\bar\theta^2 =\bar\theta^1 \hat P_0$ where $\hat P_0= C^{-1} P^T_0 C$.
Since
$\delta\theta^i$ satisfies the same conditions as $\theta$
equation (3.4) implies  that
$$
\eqalign{
\gamma^{\mu}+\hat P_0\gamma^\mu P_0 &=0,  \ \mu\ \in \ D_0,\cr
\gamma^\mu-\hat P_0\gamma^\mu P_0 &=0,  \ \mu\ \in \ N_0,}
\eqn\bccon
$$
Without loss of generality we can choose $N_0=\{0,1,2,\dots,p\}$ for one
D-brane. Taking
$P_0=\gamma^ {01\dots p}$ and using the properties  $\hat P_0=C^{-1}
P^T_0 C=(-1)^{p}P_0^{-1}$ and $P_0^{-1}=-(-1)^
{{1\over2}p(p+1)}P_0$, we find that the boundary conditions \bccon\ are
satisfied. We must now return to the $\delta X^{\mu}$ terms
in \boundarytwo. Clearly these vanish for Dirichlet boundary conditions.
For Neumann boundary conditions we require that
$\partial_\s\theta^2 = -P_0\partial_\s\theta^1$ 
and $\partial_\s\theta^2 = -P_\pi\partial_\s\theta^1$ at the two end points
$\s=0$ and $\s=\pi$ respectively. However this 
follows from the observation that the equations of motion \fulleqofm\ 
for $\theta^1$ and $\theta^2$ are related by a reflection 
$\s\leftrightarrow -\s$, (this is easily seen in the conformal gauge 
$g_{\alpha\beta}=e^\phi\eta_{\alpha\beta}$). It is now possible to see
that all terms in \boundarytwo\ vanish, including the terms cubic in 
$\theta$.

In IIA theory string theory, we choose $\theta^1$ and $\theta^2$ to have
opposite chiralities. Since $P_0\gamma^{11}=(-1)^{(p+1)(D-p-1)}\gamma^{11}P_0$
we find that the condition $\theta^1=P_0\theta^2$ requires D-branes to
have even $p$. In type IIB string theory on the other hand, $\theta^1$
and $\theta^2$
have the same chirality and we therefore find D-branes have odd $p$.
To summarise, if we have an open string which goes from one  D-brane
to another or from one D-brane to itself the $\theta^i$ are subject
to the constraints
$$
\theta^2=P_0\theta^1,\ \ \theta^2=P_\pi\theta^1,
$$
at $\sigma=0$ and $\sigma=\pi$ respectively.
In general if the
D-brane lies in 
some other directions then $P_0$ is given by the product of all the
$\gamma$-matrices in the longitudinal directions of the D-brane. The
matrix $P_\pi$ is formed in the same way.

Next we must consider the effect of the boundary conditions \bccon, which will
break some of the supersymmetries.  
Clearly the 
only supersymmetry variations $\delta\theta^i$ that preserve the boundary 
conditions on $\theta^i$ are those for which
$$
\epsilon^2 = P_0\epsilon^1\ ,\quad \epsilon^1 = P_0^{-1}P_\pi \epsilon^1\ .
\eqn\econ
$$
Now consider the $X^{\mu}$ boundary conditions at $\sigma=0$. One finds
$$\eqalign{
\delta X^{\mu}\mid_{\sigma=0} &= i\bar\epsilon^1
\left[\gamma^{\mu} + \hat P_0 \gamma^{\mu}P_0\right]
\theta^1\mid_{\sigma=0}\ ,\cr
\delta \partial_\sigma X^{\mu}\mid_{\sigma=0} &= i\bar\epsilon^1
\left[\gamma^{\mu} - \hat P_0 \gamma^{\mu}P_0\right]
\partial_\sigma \theta^1\mid_{\sigma=0}\ .\cr
}
$$
Demanding that $\delta X^\mu=0$ for $\mu\in D_0$ and 
$\delta\partial_\s X^\mu=0$ for $\mu\in N_0$ leads to
\bccon. Analogous conditions arise for $\sigma = \pi$. 
Hence we learn that the boundary conditions are compatible with
supersymmetry

In fact we see that the boundary conditions for the scalars are determined by
the supersymmetry projectors \econ. For example 
one can readily see that if 
$P_0 = \gamma^{012...p}$ then to preserve supersymmetry we must impose 
Neumann boundary conditions on $X^\mu$ 
for $\mu = 0,...,p$ and Dirichlet boundary conditions for 
$\mu=p+1,...,9$. Thus by requiring a that a particular spacetime
supersymmetry \econ\ is preserved we learn that the $\s=0$
end point of the string is fixed to lie on a surface parallel to the
$(x^0,x^1,x^2,...,x^p)$ plane.
Similarly one can determine on which plane
the $\sigma=\pi$ end point of the open string must end from a knowledge of
$P_\pi$. 
Therefore we could define a D-brane as an open string in type II string theory
that preserves the spacetime supersymmetries \econ. 
It then follows as a corollary
that a D$p$-brane with sixteen supersymmetries (i.e. with $P_0=P_\p$) 
is $p$-dimensional spacelike plane on which open strings can end [\DLP]. 

There is another form of boundary condition that we want to consider here. 
Namely if we 
suspend a D-brane configuration between two NS-Fivebranes. This was
first considered in [\HW] where D-threebranes were placed between 
NS-fivebranes. This has the effect of dimensionally reducing the
worldvolume of the intersection. 
In addition  boundary conditions must be imposed where the D-branes
meet the NS-fivebranes and these can
remove massless modes [\HW]. 

To be concrete consider two NS-fivebranes 
parallel to the $(x^0,x^1,x^2,x^3,x^4,x^5)$ plane which  
have been introduced into a brane configuration,  resulting in  a 
particular direction, say $x^6$, being reduced to an interval. 
The presence of the NS-fivebranes
breaks half of the remaining supersymmetries, namely those for which
$$
\epsilon^1 = \gamma^{012345}\epsilon \equiv P_{NS}\epsilon^1\ .
\eqn\NSpro
$$
We want to find out which string states are compatible with the remaining
supersymmetries. Now the end points of the string are constrained to lie on an
interval for all time. Since their motion is determined from the Green-Schwarz
action it follows that they can not carry any momentum along $x^6$. Therefore
the string end points must  be fixed at a definite value of $x^6$. 
Preserving supersymmetry requires that
$$
\delta X^6\mid_{\s=0} = \bar\epsilon^1[\gamma^6 + \hat P_0\gamma^6P_0]
\theta^1
=0\ ,
\eqn\con
$$
at $\s=0$.
There is  an analogous condition at $\s=\p$ but since 
$P_0\theta^1 = P_\p \theta^1$ and 
$P_0\epsilon^1 = P_\p \epsilon^1$ this yields an identical  
constraint. 
One can see that \con\ vanishes automatically if $X^6$ has
DD, DN or ND boundary conditions. However, if the boundary conditions for $X^6$
are NN then we find that to preserve the spacetime supersymmetry we must have
$$
\bar\epsilon^1\gamma^6 \theta^1|_{\s=0,\p}=0\ ,
\eqn\NSsusy
$$
for all $\epsilon^1$ such that 
$P_{NS}\epsilon^1=P_0^{-1}P_\pi\epsilon^1=\epsilon^1$.
Next we write $\epsilon^1 = {1\over4}(1+P_0^{-1}P_\pi)(1+P_{NS})\eta$ 
for an arbitrary $\eta$ and  
substitute this for $\epsilon^1$ in \NSsusy. 
Since \NSsusy\ is now true for all $\eta$
we find that the boundary conditions introduced by this dimensional reduction
must project out  spinors
$$ 
(1+P_\p^{-1}P_0)(1 - P_{NS}) \theta^1|_{\s=0,\p}=0\ .
$$
If we further assume, as we will in what follows,  
that $P_\p$, $P_0$ and $P_{NS}$ all commute then this simply becomes, 
$$
(1-P_{NS})\theta^1|_{\s=0,\p}=0\ ,
\eqn\classcon
$$
where we have used the fact that $P^{-1}_\p P_0\theta^1=\theta^1$ at $\s=0,\p$.
There are other ways to compactify a brane configuration 
which  reduce the massless spectrum. For example one could place the 
configuration between two D-fivebranes [\HW] or
compactify $x^6$ on the orbifold $S^1/{\bf Z}_2$. 
The analysis of these situations
follows in an analogous manner.

Our final step in analysing the physical spectrum of states is 
to go to the light cone gauge.
In this case the equations of motion simplify considerably.
Since the theory is invariant under worldsheet reparameterisations
we  choose the conformal 
gauge $ g_{\alpha \beta }= \eta_{\alpha \beta}e^\phi$.  
We may also use the residual conformal symmetry to choose
$$
X^+ \equiv X^0+X^1=x^+ + 2\alpha'p^+ \tau ,
$$
where the normalisation of $p^+$ is appropriate for an open string.
Using the $\kappa$-symmetry of equation \ksym\ 
we may make one further gauge choice $\gamma^{01}\theta^i = \theta^i$ 
and reduce our spinor to
$$
S^i = {1\over2}(1+\gamma^{01})\theta^i\ .
$$
Note that the light cone projector involves a choice of direction (here we 
assume it is $x^1$) and sign (here we assumed a plus sign). However these
can not be chosen arbitrarily since   
$\gamma^{01}$ must commute with the other
projectors $P^{-1}P_\p$ and $P_{NS}$. 
In addition, in some two-dimensional cases these  projectors 
may imply $\gamma^{01}\theta^i=\pm \theta^i$.
This means that the $\theta^i$ are chiral fermions on the worldsheet. 
However we
must chose the light cone projector to act non-trivially, 
otherwise the unphysical field components will
not be projected out. 
Thus in these cases we must chose the light cone condition 
$\gamma^{01}\theta^i = \mp\theta^i$. We will assume these conditions
are satisfied in what follows.

With these choices, the first of the equations \fulleqofm\ can be
used to solve for $X^- $ and $P^-$. The equations
of motion for the remaining variables take on the very simple form
$$\eqalign{
(\partial^2_\s-\partial_\t^2)X^\mu&=0\ ,\cr
(\partial_\s+\partial_\t)\theta^1&=0\ ,\cr
(\partial_\s-\partial_\t)\theta^2&=0\ ,\cr}
\eqn\eqofm
$$
where here, and for the rest of this paper, $\mu\ne 0,1$.
Let us first  solve for the scalars fields $X^\mu$.
There are four 
possible combinations of boundary conditions for an open string: 
NN, DD, ND and DN whose expansions are 
$$
\eqalign{
X^{\mu} &= x^{\mu} + 2\alpha'\alpha_0^\mu\tau 
+ i\sqrt{\alpha'}\sum_n{\alpha_n^\mu\over n}e^{-in\tau}
{\rm cos} n\sigma\ ,\cr
X^{\mu} &= x^{\mu} + (y^\mu-x^\mu){\s\over\p}
+\sqrt{\alpha'}\sum_n{\alpha_n^\mu\over n}e^{-in\tau}
{\rm sin} n\sigma \ ,\cr
X^{\mu} &=  y^{\mu} + i\sqrt{\alpha'}\sum_r{\alpha_r^\mu\over r}e^{-in\tau}
{\rm cos} r\sigma\ ,\cr
X^{\mu} &=  x^{\mu} + \sqrt{\alpha'}\sum_r{\alpha_r^\mu\over r}e^{-in\tau}
{\rm sin} r\sigma\ ,\cr
}
\eqn\Xexpansion
$$
respectively. In \Xexpansion\ and what follows  
$n \in {\bf Z}$ and $r \in {\bf Z}+{1\over2}$. Note that
there is only momentum in directions with NN boundary conditions.

Now we consider the Spinor fields $S^i$ which, as a result of \eqofm\ 
take the form
$$\eqalign{
S^1 = {1\over\sqrt{2}}\sum_n S_n^1e^{-in(\tau-\s)}
+ {1\over\sqrt{2}}\sum_r S_r^1e^{-in(\tau-\s)}\ ,\cr
S^2 = {1\over\sqrt{2}}\sum_n S_n^2e^{-in(\tau+\s)}
+ {1\over\sqrt{2}}\sum_r S_r^2e^{-in(\tau+\s)}\ .\cr
}
$$
Next we must impose the boundary conditions $S^2 = P_0 S^1$ at $\s=0$ and
$S^2 = P_\p S^1$ at $\s=\p$. This leads to 
$$
\left.\matrix{S_n^2=P_0 S_n^1\cr 
              S_r^2=P_0 S_r^1\cr}\right\}
{\ {\rm at}\ \s=0\ ,}\quad\quad
\left.\matrix{S_n^2=P_\p S_n^1\cr 
              S_r^2=-P_\p S_r^1\cr}\right\}
{\ {\rm at}\ \s=\p\ .}
$$
Clearly two of these equations simply determine $S^2_n$ and $S^2_r$ in
terms of $S^1_n$ and $S^1_r$. However the other two equations lead to the
constraints
$$
S^1_n = P_0^{-1}P_\p S^1_n\ ,\quad S^1_r = -P_0^{-1}P_\p S^1_r\ .
\eqn\SNR
$$

In the quantum theory $X^\mu$ and $S^1$ are promoted to operators which obey
the standard commutation and anti-commutations relations respectively. We 
introduce a vacuum state $|0>$ which is annihilated by all the modes
in $X^\mu$ and $S^1$ with positive frequency; $n,r > 0$. 
Due to normal ordering ambiguities 
the zero point energy (i.e. the rest mass squared of $|0>$) 
is given by the ``intercept'' 
$a =-(N(\alpha_n) - N(S^1_n))/24+(N(\alpha_r) - N(S^1_r))/48$, where, for 
example, $N(\alpha_n)$ 
counts the number of bosonic integer moded oscillators [\BN]. The action of
a non-negatively moded oscillator on the ground state then raises the mass
level by $n$ or $r$. 
Thus the ground state is degenerate as $S^1_0$ and $\alpha^\mu_0$
both map $|0>$ to itself. 
In fact the action of $\alpha^\mu_0$ simply generates 
spacetime momentum so we will ignore it here. 

Let us suppose that 
we have $N$ $S_0^{1A}$ oscillators $A=1,...,N$ 
(and we assume that $N$ is even). On behalf of the 
anti-commutation relation $\{S^{1A}_0,S_0^{1A}\}=\delta^{AB}$, the $S_0^{1A}$
form a Clifford algebra in $N$-dimensions acting on the ground state.
There is a unique non-trivial representation of this algebra which can be  
constructed as follows. We 
start by defining $N/2$ helicity raising and $N/2$ helicity 
lowering operators in the usual manor. 
Then, from the lowest helicity component  
$|0:l>$ of $|0>$ (which is annihilated by all the lowering operators), 
we obtain $2^{N/2}$ independent states by acting with
the $N/2$ distinct raising operators. These create equal numbers of bosonic
and fermionic states which fill out a
supersymmetric multiplet. Note that a spacetime supersymmetry algebra with
$2N$ charges splits into $N$ ``null'' supersymmetries with $\{Q,Q\}=0$
and $N$ ``physical'' supersymmetries with $\{Q,Q\}=1$. The ``null''
supersymmetries act trivially on the spectrum 
and obey $\gamma^{01}Q=-Q$ if we choose spacelike momentum in the $x^1$
direction. The ``physical''
supersymmetries are the ones used above to create the non-trivial 
representation of a Clifford algebra in $N$ dimensions. In
general the states obtained by this method are not CPT self-conjugate and
so, according to the CPT theorem, we must also include their CPT 
conjugates to obtain
a physically acceptable representation. However, for the special maximal 
case of  sixteen supersymmetries, the vector multiplet is CPT self-conjugate.

We must also decide what we mean by the constraint 
\classcon\ in the quantum theory.
If we assume that \classcon\ acts directly on 
the fields $S^1$ then we run into
severe problems. For in this case imposing \classcon\ reduces 
the number of independent spinors $S^{1}$ and therefore changes
the intercept. The resulting spectrum would then have different masses
(quite possibly tachyonic) and could not be viewed as a subset of the original
multiplet. Thus we must interpret \classcon\ as applying to 
the action of $S^{1}$ on the ground state $|0>$. In particular we find
for the zero modes that
$$
(1-P_{NS})S^1_0|0>=0\ .
\eqn\sixconstraint
$$

In summary  the spinors $S^2_n$ and $S^2_r$ have been eliminated 
in terms of $S^1_n$ and $S_r^1$ which are subject to \SNR. 
If we include NS-fivebranes then the spectrum of massless states is
also subjected to the constraint  \sixconstraint. 
Note that, since $(P_0^{-1}P_\p )^2=1$ and ${\rm Tr}(P_0^{-1}P_\p ) = 0$
unless $P_0^{-1}P_\p$ is a multiple
of the identity, there
are only three relevant cases for $P_0^{-1}P_\p$: 
(1)  all eight eigenvalues
equal one, (2) all eight eigenvalues equal minus one, (3) 
four eigenvalues are plus one and four eigenvalues are minus one.
Let us consider in detail these three cases.

First we consider case (1), corresponding to  
parallel D$p$-branes with $P_0=P_\p$.   
We see immediately that $S^1_r=S^2_r=0$ and $S^1_n$
is unconstrained. This leaves us with eight independent integer moded spinors
and sixteen spacetime supersymmetries.
The fields $X^\mu$ have either DD or NN boundary conditions 
and hence there are
also eight independent integer moded scalars. This leads to a 
Clifford algebra in eight dimensions  
represented by a sixteen-dimensional massless ground state. 
Finally, if there are $N$ D$p$-branes, then we assign labels
to the string end points to indicate which D$p$-brane they lie on. 
The states of the quantised string then come in an adjoint
representation of $U(N)$ and, since it is a sixteen-dimensional, 
it follows that the ground state is a maximal super-Yang-Mills vector 
multiplet in $p+1$ dimensions [\Witten].

If we suspend these D-branes between two NS-fivebranes 
then we must include the 
projection \sixconstraint. This  removes 
half of the sixteen spacetime 
supersymmetries and leaves only four independent $S_0^{1A}$ acting on the 
ground state. 
Applying the remaining raising operators to 
$|0:l>$ produces a vector
multiplet with eight supersymmetries. In fact since there are now only
four integer moded spinors, 
and two helicity raising operators, we only obtain
states with helicities $-1,-1/2,0$. However the CPT theorem asserts that
we must add in the positive helicity states to obtain a complete
vector multiplet with eight supersymmetries.
This is the situation in [\HW] for
D-threebranes suspended between two NS-fivebranes.

Next we consider case (2), corresponding to 
an open string with one end on a D$p$-brane and the 
other on an anti-D$p$-brane. Here $P_0=-P_\p$ and  
no spacetime supersymmetries are preserved.  
The corresponding $X^\mu$ fields are the same as the 
$P_0=P_\p$ case above. Therefore there are eight integer
moded scalars but now there are eight half-integer moded fermions. 
The ground state is a Clifford algebra in zero dimensions and
is one-dimensional.
However the intercept is $-1/2$ so the ground state is a tachyon [\Green]. 

Finally we consider case (3). First we consider a degenerate situation
where $P_0^{-1}P_\p = -\gamma^{23456789}$. Here 
$P_0^{-1}P_\p S^1 = -S^1$ and hence there are no $S_n^1$ and eight $S_r^1$.
This corresponds to eight
$X^{\mu}$ with DN or ND boundary conditions (i.e. an open string stretched
between a D$p$-brane and a D$(p+8)$-brane). 
Thus the gound state is massless, forms a one-dimensional
Clifford algebra in zero dimensions and is a singlet
under spacetime supersymmetry [\PW]. 
  
In the general situation of case (3) there are
four integer moded fermions $S^1_n$ and four half-integer moded fermions
$S^1_r$. In addition there are 
four $X^\mu$ with either
DD or NN boundary conditions and four $X^\mu$ with ND or DN 
boundary conditions (i.e an open string streched between a D$p$-brane and
a D$(p+4)$-brane). Thus  the ground 
state forms a four-dimensional representation of 
Clifford algebra in four dimensions.
If there are $N_1$ D-branes corresponding to $P_0$ and $N_2$ D-branes
corresponding to $P_\pi$ then, after labelling each end point, 
the states of this string appear in the
bi-fundamental representation of $U(N_1)\times U(N_2)$. Since the states
are not in the adjoint representation, they must form a hyper multiplet, 
dimensionally reduced to the intersection of the D-branes [\BSV]. 
The four states that we obtain from quantising this
string are not CPT invariant. 
To cure this we note that these open strings are unorientated and so we must 
also include strings with $P_\pi$ and $P_0$ applied at the end points
$\s=0$ and $\s=\pi$ respectively. This adds another copy of the above 
states, resulting in a CPT invariant mutliplet.

Let us suspend this D-brane configuration between two NS-fivebranes. We
must then include the effect of \sixconstraint\  on the zero-modes.  
In general this leaves two non-vanishing zero modes on $|0:l>$ giving a 
Wess-Zumino multiplet. After including the CPT conjugate states we 
therefore obtain two bosons and two fermions.
An example of such a configuration is
to chose  
$P_0 = \gamma^{0123456}$, $P_\pi=\gamma^{0123678}$ and rather than 
adding NS-fivebranes we compactify 
$x^6$ on a the orbifold $S^1/{\bf Z}_2 \cong [0,1]$. This corresponds to two 
D-sixbranes
intersecting over a four-brane preserving $N=1$ supersymmetry in four 
dimensions
$$
\matrix{D6:&1&2&3&4&5&6& & \cr
        D6:&1&2&3& & &6&7&8\cr}
$$
where $x^6 \in [0,1]$. After compactification the 
surviving supersymmetries satisfy 
$\gamma^6\epsilon = \epsilon$. In terms of the chiral components this gives 
$\epsilon^1 = \gamma^6\epsilon^2 = \gamma^{012345}\epsilon^1$. This is 
the same projector as \NSpro\ and therefore we  find the same
constraint \sixconstraint\ on $S_0^1$.
The ground state forms  a  chiral 
four-dimensional $N=1$ Wess-Zumino multiplet in the bi-fundamental of
the two D-sixbrane gauge groups.
 
Finally, an interesting degenerate case occurs if 
$P^{-1}_0 P_\p S_0^1 = -P_{NS}S^1_0$ 
then, since $P_{NS}S_0^1 = -P^{-1}_0 P_\p S_0^1 = -S_0^1$, 
we find $S_0^1|0:l>=0$.
Thus there is no spacetime supersymmetry in such a string and the only
surviving massless modes are the lowest helicity components of the ground 
state. Since the ground state is a hyper multiplet this gives two helicity
$-{1\over2}$ fermions.
To be specific one could take $P_0=\gamma^{01236}$,  
$P_\p=\gamma^{01346}$ and $P_{NS}= \gamma^{012345}$ which can be pictured as
$$
\matrix{NS5:&0&1&2&3&4&5& \cr
         D4:&0&1&2&3& & &6\cr
         D4:&0&1& & &4&5&6\cr}
$$
i.e. two D-fourbranes intersecting over a twobrane and suspended between two
NS-fivebranes. 
After performing the dimensional reduction induced by the NS-fivebranes 
one finds two chiral fermions in the resulting two-dimensional
theory. 

Clearly there are other possibilities which we have not analysed here. 
For example one could  add more  
NS-fivebranes, as in  brane box
constructions, dimensionally reducing
additional directions of a brane configuration. 

\refout

\end